# Novel Concept of the Magmatic Heat Extraction

Mark Labinov, PhD, PE, CEM
Principal, Thermal/ Power, +4Pi
College of Engineering, Technology and Architecture, University of Hartford,
200 Bloomfiled Ave, West Hartford, CT, USA, 06117
E-mail: labinovmark11@gmail.com

**ABSTRACT**

Enhanced geothermal systems, EGS are the primary sources of interest nowadays and for good reason-they have the highest possible thermal potential through the perspective of accessing the magmatic heat. Magma is a complex, high-temperature fluidic substance. Temperatures of most magmas are in the range between 700 C and 1300 C (1300 °F and 2400 °F), In the rare occasions carbonatite melts may be as cool as 600 °C, and komatite melts may have been as hot as 1650°C. [1]. The perspective to access this heat on a permanent basis is a primal EGS challenge. At the same time it is the least accessible geothermal source among the listed perspective EGS- projects. That's associated with the traditional ways of that access; usage of the heat exchanger with recirculating gas as a working fluid, followed by the Brayton cycle, similar to the older gas nuclear power plant design, whereas instead of the nuclear reaction there's the magma pool [2,3] The paper presents a novel concept based on the Retrograde Condensation phenomenon; it presents its potential benefits and Energy Return On Investment (EROI) forecasts.

*1. Introduction and Background*

'Tapping energy stored in the molten rocks beneath the surface of the Earth had been a dream for a very long time. Geologists and laymen, looking at the lava flowing from the active volcano have long wished that this energy could be harnessed There is no doubt that this subsurface reserves of heat energy are enormous' [2]. MIT EGS assessment for the US usage report [1] states:

'The accessible geothermal resource, based on existing extractive technology, is large and contained in a continuum of grades ranging from today's hydrothermal, convective systems through high-and mid-grade EGS resources (located primarily in the western United States) to the very large, conduction-dominated contributions in the deep basement and sedimentary rock formations throughout the country. By evaluating an extensive database of bottom-hole temperature and regional geologic data (rock types, stress levels, surface temperatures, etc.), we have estimated the total EGS resource base to be more than 13 million exajoules ($1J=10^{-18}EJ$). Using reasonable assumptions regarding how heat would be mined from stimulated EGS reservoirs, we also estimated the extractable portion to exceed 200,000 EJ or about 2,000 times the annual consumption of primary energy in the United States in 2005. With technology improvements, the economically extractable amount of useful energy could increase by a factor of 10 or more, thus making EGS sustainable for centuries.'

The approaches to extracting of the EGS heat from the ground had been so far traditional: heat mining involved drilling, heat exchangers and a thermodynamic cycle for the power delivery. Most of the geothermal sources, especially magmatic sources were treated as ground–based nuclear reactors- the first section of the bi-sectioned power plant. The purpose of this paper is to introduce the novel concept of extracting that energy, based on the defined retrograde condensation (RC) phenomenon in the two-phase zone [4] The paper describes the phenomenon in question, suggests the geothermal fluidic system and makes a first attempt to apply the EROI approach to prove the advantage of such system over the traditional geothermal approaches to EGS.

The RC -phenomenon takes place in the two-phase zone of the pure fluid although the actual definition comes from the binary mixture domain [5]. It manifests itself clearly in the medieval 'cannon experiment' which is considered an origin of the 'power of steam'. (Fig.1). A cannon was filled with water to the brim, then-sealed shut and stuck into the burning fire.

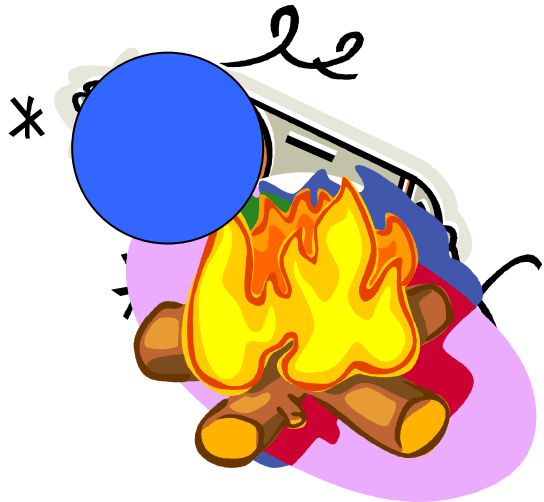

Fig.1. The cannon was filled with water, sealed and-heated until it ruptured.

Eventually the cannon ruptured open and there was a lot of steam coming out. The conclusion was that 'steam had a lot of power'; that eventually led to steam engines, Watt's locomotive and modern power plants. Although the conclusion was useful, the explanation was wrong. It was superheated water that ruptured the cannon. Fig.2 shows the process thermodynamically.

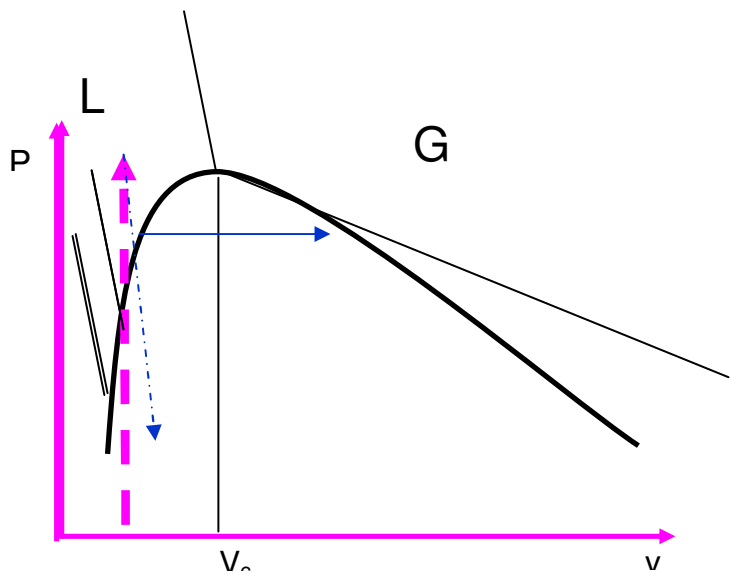

Fig.2 Thermodynamic picture of the cannon experiment; L-liquid, G-gas.

The process of the constant-volume heating when the cannon is heated to the left of the critical specific volume $v_c$ is as follows: steam quality x rises at first but after reaching a certain level it starts to drop, the steam bubble collapses-and when the saturation line is crossed-we reach the superheated liquid, where pressure rises exponentially. When the rupture takes place, the isentropic pressure drop occurs instantaneously; the fluid reaches the saturation line with the subsequent flash boiling. That would be the steam emerging en masse, the cloud people saw.

In case the cannon is filled only up to ¼ or so the process happens to the right of the critical specific volume. Water evaporates fully into steam and the pressure of the superheated steam rises linearly with the temperature; even when the cannon volume reaches the flame temperature, there would be not enough pressure for a rupture.

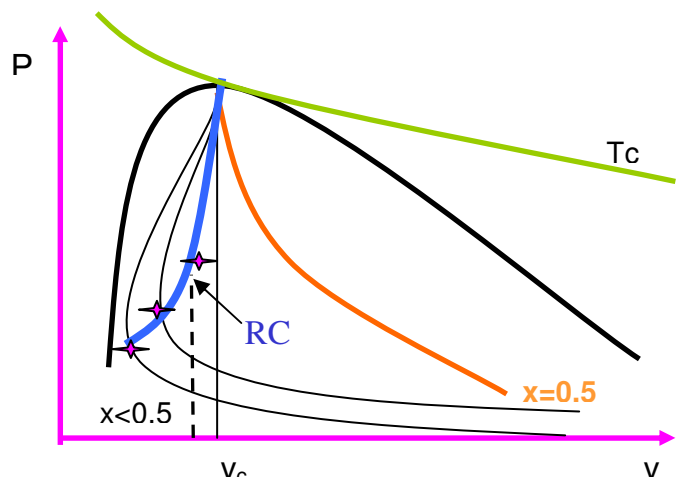

Fig.3 Retrograde Condensation (RC) Curve

Thus for every specific volume lower than the critical there must be a point on the constant-volume line where quality rate changes its sign from positive to negative. That point is designated as a Retrograde Condensation (RC) point and the locus of these points to the left of the critical volume as presented on Fig 3 is called the RC-curve per equation (1):

$$x_{RC} = \frac{dv_l/dT}{[dv_l/dT - dv_{vp}/dT]} \quad (1)$$

As a conglomerate of points where quality rate changes its sign while external influence (heat input) stays the same, the RC-curve defines the borderline between the stable and unstable states of thermodynamic equilibrium. That would indicate a necessary condition for the self-sustaining quality/evaporation/condensation oscillations around each such location. The analytical solution of the mass transfer equation in the vicinity of the RC-point in the near-critical region of the fluid resulted in the self-sustaining limit-cycle type steam quality oscillations and thus–pressure [4]:

$$\sqrt{\frac{x}{(1-x)}} \cong \sin[\omega\tau + \sin^{-1}\sqrt{\frac{x_{rc}}{1-x_{Rc}}}] \quad (2)$$

$$\omega = f(T_c, T_s, physical\ properties\ of\ substance)$$

The equation (2) lays a foundation for the novel type of thermodynamic machinery, based on self-sustaining oscillations with frequency ω depending not on the geometry but on the RC-location and physical properties o the working fluid. Non-equilibrium pressure oscillates in a limit cycle around the RC-point with the saturation pressure as an average value and amplitude defined by the quality oscillations (2).

## 3. Description: Novel Concept for the Geothermal Power Generation

*3.1 EROI approach to the geothermal systems; choosing a perspective one.*

The primary disadvantage of the traditional EGS heat extraction systems is their complexity [6]. References [3,4] list the challenges associated with the magmatic heat extraction with Brayton cycle as a power delivery branch (Sandia approach). As a part of the research project Sandia constructed the magma simulation facility with the goal to check the viability of the insertion of the heat exchanger into the magma pool. The heat extraction required not only then novel drilling techniques for the source tapping but also-an advanced long-term heat exchanger design to connect with the power delivery branch. Traditional system thus deals with at least two heat transfer systems, one of which, the power delivery branch is bounded in its efficiency by the Kurzon-Ahlborn formulae [6]:

$$\eta = 1 - \sqrt{\frac{T_c}{T_h}} \qquad (3)$$

Here $T_c$, K- the lowest temperature in the thermodynamic cycle of the delivery branch, $T_h$, K the highest. Per the energy transformation devices' classification presented in [9] this arrangement corresponds to the case of two energy transformers over one path, including all the traditional ones: internal consumption, external energy sources and an indirect energy supply piece. Per equation (3) the energy transformation in the delivery branch is severely limited due to thermodynamics. That becomes especially important when considering the magmatic environment; traditional system in this case faces the energy return on investment (EROI) - limitations. EROI is defined as follows [8]:

$$\frac{El.\,energy\ produced\ over\ life\ cycle-}{El.\,energy\ (building\ maintenance, decomission)} \quad (4)$$

EROI- approach is currently experiencing the revival of interest as a non-ambiguous, reliable way to access the new power generation projects, also maintenance and decommission of the older plants. Per [10], wind and solar power look increasingly promising while geothermal still lagging in due to the significant drilling energy expenses, and because of the inconsistency of the geothermal sources especially when it comes to EGS [1]. Hall and others in [11] defined a minimum sustainable EROI-requirement as 3:1. Mansure in [8] applies the EROI theory to the geothermal energy extraction and comes up with the corrected 'closed loop' formulae:

$$EROI_2 = \frac{EROI_1 - f}{(1-f)} \quad (5)$$

Here $EROI_1$ is the EROI from (4), f- is the fractional reduction of the necessary input electric energy ,due to 'closing of the loop'-the substitution of it by the usage of the electricity from the geothermal power output. In short, for the geothermal power to compete, he encourages reinvestment of its output to compensate for the electrical input generated by external non-renewable means. He logically specifies that such arrangement within the conventional geothermal domain would mean the redirection of the fraction of the output to replace the maintenance input and to participate in the new geothermal plants' construction.

This leads to the criteria that among the geothermal designs the most perspective one should be the most enabling one to the 'closing of the loop' arrangement (Fig.4).

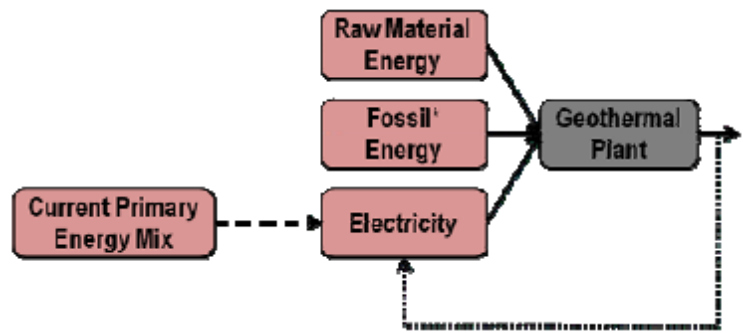

Fig.4 Closing the loop [8]

## 3.2. The RC- based arrangement of the magma- based power generation

The point of the new arrangement is to have a permanent prime mover located in the magma chamber per Fig 5; this gets rid of the secondary power generation branch and simplifies the design considerably.

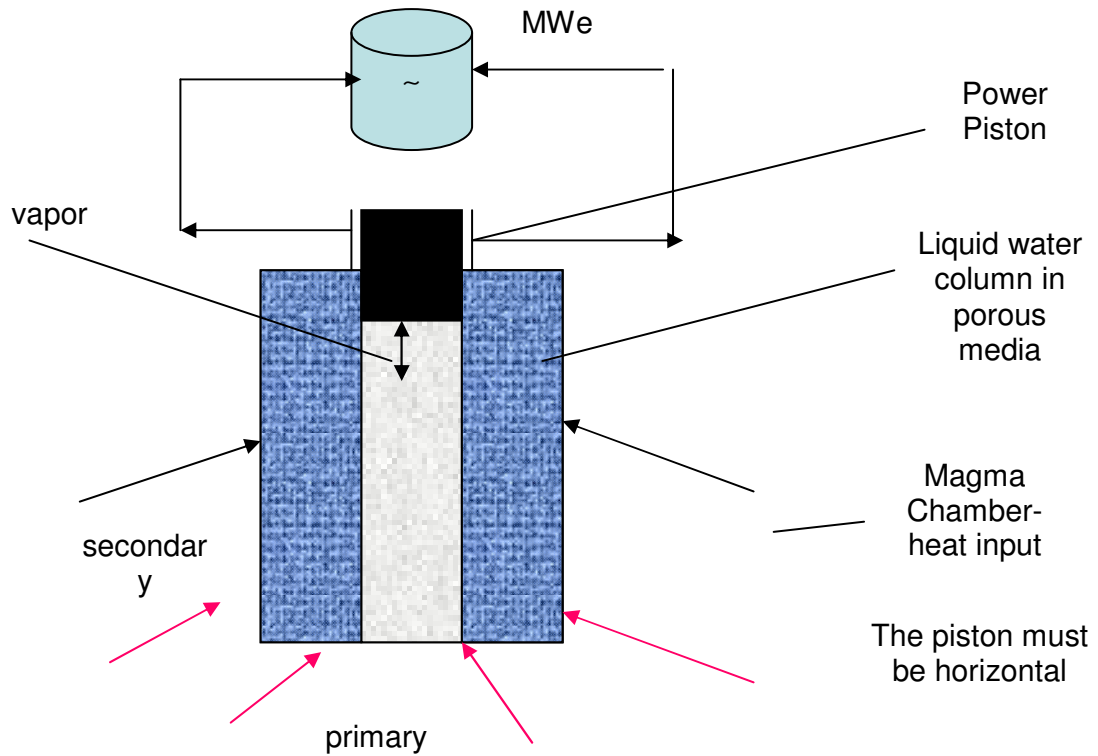

Fig.5 The Fluidyne-type [12] power generation assembly for the magmatic heat extraction

The assembly consists of the casing, filled with porous media and a free volume in its center. In the working system this volume is filled with vapor which is in equilibrium with the liquid filling the porous media- both volumes heated first through a special valve–controlled process to the RC point defined and Ps- average pressure. When excited, at the RC-point the system oscillates, delivering power output encountering thermal losses. The main difference when compared to the steam engine is the closed assembly here, while for example, in the traditional Rankin arrangement in the power plants we have a water loop, needing additional equipment-turbines, condensers, superheaters, pumps and cooling towers. The presented arrangement more resembles a hypothetical situation when the vapor produced in the nuclear reactor is used directly to generate power, which never happens in the real nuclear plants even in the ones with an embedded power delivery branch.

*3.3 The EROI-advantages of the novel magmatic arrangement*

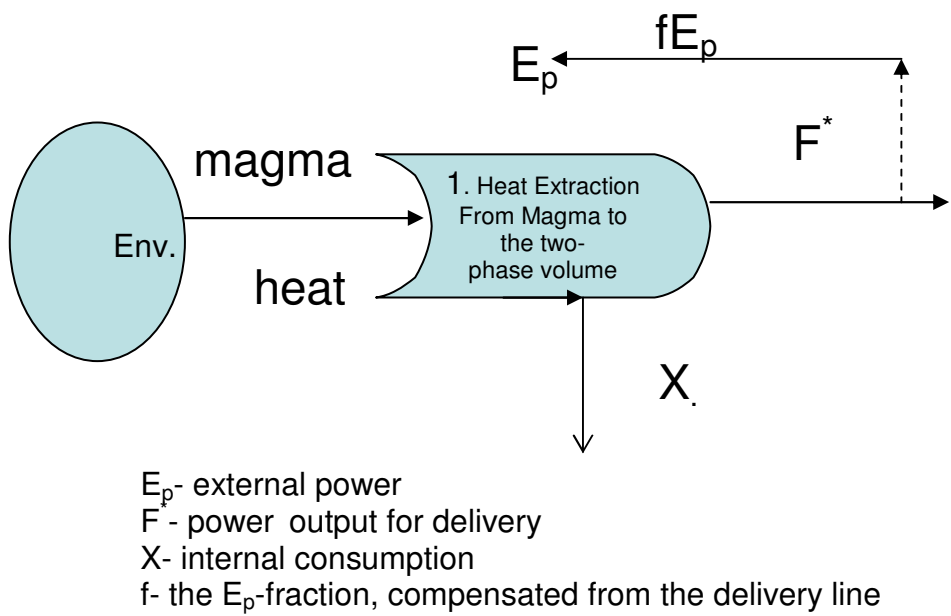

$E_p$- external power
$F^*$- power output for delivery
X- internal consumption
f- the $E_p$-fraction, compensated from the delivery line

Fig. 6 The novel arrangement: system energy flows and definitions [9]

Fig. 6 presents the novel arrangement as an energy transformation system per the methodology in [9] Per Mansure's definitions (5) the EROIs without the closed loop and with the loopfor this arrangement are defined as follows:

$$\boxed{\begin{aligned} EROI_{1-s} &= F^* / (X + E_p) \\ EROI_{2-s} &= \frac{F^* - fE_p}{X + E_p(1-f)} = \\ &\quad \frac{F^*/E_p - f}{X/E_p + (1-f)} \end{aligned}} \quad (6)$$

In case the external power fraction $fE_p$ is compensated by the usage of the system power output, the remaining external power needed becomes $E_p(1-f)$.

The novel power generation schematic on the Fig.6, built according to the principles of [9] point out to the following crucial advantages per (6):

- Simplicity

- High thermal potential of the heat losses- suitable for the internal utilization

- Implicit availability of the compensating for nearly all external power through the closing of the loop; it would be proper to seek the most precise optimum here because in the conditions of the volcano self-sustainability and self-reliance of the plant become a must.

An optimization analysis of the Fig. 6 system at the constant ratio of $X/E_p=1$ reveals $EROI_{2-s}$ values in the middle 30s with high relative levels of the power generated. That puts the magmatic arrangement on the line with the disturbing technology innovations in the area of the renewable energy.

## 4. Conclusion

The paper offers a novel, perspective approach to the EGS- extraction of the magmatic heat by the usage of the RC-thermodynamic phenomena. It is the intention of the author to follow through with the first–order concept- process model, identifying system design issues and specific challenges.

## Nomenclature

$EROI$ – energy return on investment
$x$ – steam quality parameter
$T$ – temperature, $K$
$T_c$ – critica.l temperature, $K$
$T_s$ – saturation temperature for the $RC - \text{point}$, $K$
$P$ – pressure, $Pa$
$P_s$ – saturation pressure for the $RC - \text{point}$, $Pa$
$\omega$ – cyclic frequency of oscillations
$v$ – specific volume, $m^3/kg$
$v_c$ – critical specific volume, $m^3/kg$